\documentclass[preprint]{elsarticle}
\usepackage{amssymb}
\usepackage{graphicx}
\usepackage{dcolumn}
\usepackage{bm}
\journal{}

\begin{document}

\title{Band structure calculations of Ti\raisebox{-.2ex}{\scriptsize 2}FeSn: a new half-metallic compound }
\author[label1]{A. Birsan}
\author[label1]{P. Palade}
\address[label1]{National Institute of Materials Physics, PO Box MG-07, Bucharest, Romania}

\begin{abstract}
Within the framework of density functional theory, the electronic structure and magnetic properties have been studied for the Ti\raisebox{-.2ex}{\scriptsize 2}FeSn full-Heusler compound. The ferromagnetic state is found to be energetically more favorable than paramagnetic and antiferromagnetic states. The spin-polarized results show that Ti\raisebox{-.2ex}{\scriptsize 2}FeSn compound has half-metallic ferromagnetic character with a total spin moment of $2 \mu_{B}$ and a band gap in the minority spin channel of  0.489 eV, at the equilibrium lattice constant a=6.342\.{A}. 

\end{abstract}


\maketitle

\section{Introduction}
The outstanding research results, from the recent years, in the field of  Heusler compounds, made the spintronics one of the emerging disciplines, which revolutionized the thriving field of information technology, especially in the context of magnetic recording \cite{Carey2008, Sakamoto2010}. Proposed by de Groot et al. in 1983, \cite{deGroot1983} the half metallic ferromagnetism characterizes Heusler materials used nowadays for spin injection devices \cite{Coey2002} which behave like semiconductor for one spin direction and like metal for the other spin direction, leading to a $100\%$ spin polarization of electrons, at the Fermi level.  Apart from spintronics, Heusler compounds with high spin polarization, high Curie temperature and simultaneously a low saturation magnetization are suitable for ultrahigh density magnetic memory storage devices \cite{Graf2011}.

These remarkable materials have been discovered in 1903 when Fritz Heusler had shown that  $Cu_{2}MnAl$ alloy behaves ferromagnetically even though none of its constituent elements is magnetic \cite{Heusler1903}. Three decades later, the general crystal structure of these exciting materials was explained to be a face centered cubic structure \cite{Bradley1934}. Often described as ternary intermetallics, the family of Heusler compounds includes two variants: half-Heusler materials $XYZ$, with 1:1:1 composition, crystallizing in the non-centrosymmetric cubic structure $C1_{b}$, with $F\bar{4}3m$ space group and full-Heusler compounds $X_{2}YZ$   with 2:1:1  stoichiometry which typically crystallize in the cubic space group $Fm\bar{3}m$, having $Cu_{2}MnAl$ ($L2_{1}$) structure as prototype. In the particular case when the $Y$ element is more electronegative than $X$, the crystal structure observed for full-Heusler compounds, has $Hg_{2}CuTi$ prototype and $F\bar{4}3m$ space group, often named inverse Heusler structure, with $X$ atoms occupying the non-equivalent 4a (0,0,0) and 4c (1/4,1/4,1/4) Wychoff positions, while $Y$ and $Z$ atoms being placed in 4b (1/2,1/2,1/2) and 4d (3/4,3/4,3/4) position, respectively \cite{Kandpal2007}.

Today, many Heusler compounds $Mn_{2}$, $Co_{2}$, $Fe_{2}$, $Cr_{2}$ or $Ti_{2}$-based are investigated for future spintronic applications \citep{Michalska2012,Li2008,KervanS2011,Bayar2011,KervanN2011,Lei2011,KervanS2011INTREM,KervanN2012INTERM,Huang2012,Wei2012,Zheng2012,Birsan2012,Ahmadian2012}. 

The $Ti_{2}FeSn$ Heusler compound was observed in experimental investigations (SEM and X-Ray diffraction), as distinct phase in the Phase Equilibria of Ti-Fe-Sn ternary system \cite{Kozakai2007}, however no further studies had been performed on this compound. In the present paper, first principle calculations of  the magnetic moments, density of states and bandstructure based on density functional theory are performed for the Heusler compound of interest studied here, $Ti_{2}FeSn$. The obtained results are compared with electronic structures and magnetic properties of other $Ti_{2}$ - Heusler compounds analyzed in literature.

\section{METHOD OF CALCULATION}
Self-consistent band structure calculations were performed using Full Potential Linearized Augmented Plane Wave (FPLAPW) method from Wien2k code \cite{Wien}. The Perdew Burke Ernzerhof (PBE)\cite{Perdew1996A,Perdew1996B} generalized gradient approximation (GGA) was employed for the exchange and correlation interaction. The muffin-tin radii ($R_{MT}$) considered for Ti, Fe and Sn were  2.35 a.u., 2.35 a.u. and 2.3 a.u. respectively. An energy threshold of -6 Ry was set between the core and valence states. The number of plane waves selected was  determined by the cut-off condition $K_{max}\*R_{MT} = 7$ ($K_{max}$ represents the maximum modulus of the reciprocal lattice vector). For Brillouin zone (BZ) integration a 46x46x46 mesh, containing 2456k irreducible k points was used within the modified tetrahedron method \cite{Blochl1994}. The energy convergence criterion was $10^{-5}$ eV and the integrated charge difference less than $10^{-4}e/a.u.^{3}$, between two successive iteration.

\section{RESULTS  AND DISCUSSIONS}
The  enthalpy of formation, calculated by subtracting the sum of the calculated equilibrium total energies for $Ti$, $Fe$ and $Sn$, with HCP, BCC and diamond cubic structure respectively, from the minimum of the total energy of $Ti_{2}FeSn$ compound, obtained at the equilibrium lattice parameter, is -0.76 eV/atom. This negative enthalpy of formation agrees with the successful preparation of $Ti_{2}FeSn$ phase within experimental investigations \cite{Kozakai2007}. 
In the Phase Equilibria of Ti-Fe-Sn ternary system, the $Ti_{2}FeSn$ compound was observed in mixture with $Fe_{2}Ti$, $Fe_{2}TiSn$, $\beta-Ti_{6}Sn_{5}$ or $Ti_{5}Sn_{3}$.  Other identified phases were $\alpha{Fe}$, $FeTi$ and $Ti_{3}Sn$. 

Decompositions of this compound, into binary or other ternary phases are  possible escape routes which must be considered. Therefore, the following changes in enthalpy were calculated  :\\
$\Delta{H}=2E_{Ti_{2}FeSn}-E_{Fe_{2}TiSn}-E_{Ti_{3}Sn}$,\\ $\Delta{H}=3E_{Ti_{2}FeSn}-E_{Fe_{2}Ti}-E_{Ti_{5}Sn_{3}}-E_{\alpha{Fe}}$\\ 
$\Delta{H}=5E_{Ti_{2}FeSn}-4E_{FeTi}-E_{Ti_{6}Sn_{5}}-E_{\alpha{Fe}}$ \\and the positive values obtained reflect unfavorable decomposition conditions. More complicated cases could have negative changes in enthalpy, though the preparing methods can influence the arrangement of the atoms in the various sites or the persistence of metastable stoichiometries.

\begin{figure}
 \begin{center}
     \includegraphics[scale=1.0]{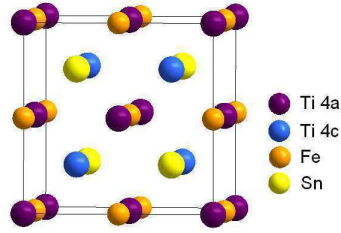}
  \end{center}
    \caption{The unit cell structure of $Ti_{2}FeSn$ compound with $Hg_{2}CuTi$  -  prototype structure.  }
        \label{fig:Ti2FeSn}  
\end{figure}

The primitive cell of $Ti_{2}FeSn$ compound  is assumed to be the cubic face centered $Hg_{2}CuTi$ prototype structure, because $Ti$ atoms are more electropositive than $Fe$ atoms \cite{Kandpal2007}. In this crystal structure, all atoms have tetrahedral symmetry $T_{d}$ and no octahedral symmetry $O_{h}$ is adopted (note that $T_{d}$ is a subgroup of $O_{h}$). $Ti$ atoms are present in two different Wyckoff positions, $Ti$-4c being located  in (0.25 0.25 0.25) while $Ti$-4a in (0 0 0), respectively, as exhibited in Fig.\ref{fig:Ti2FeSn}. This is consistent with some other Heusler compounds studied in literature, with $X$ atoms more electropositive than $Y$, which had been reported to exhibit inverse Heusler structure ($Hg_{2}CuTi$ prototype)\cite{KervanS2011,Bayar2011,KervanN2011,Lei2011,Huang2012,Wei2012,Zheng2012}. 
The lattice parameter of  $Ti_{2}FeSn$ compound is not reported in any experiment, therefore a structural optimization, within FLAPW scheme was performed with non-magnetic, antiferromagnetic and spin polarized ferromagnetic configurations. The unit cell of $Ti_{2}FeSn$ used for these calculations consists in a supercell resulted from spreading the crystal along the axis of $a$ lattice parameter. The construction of this structure was essential for antiferromagnetic configuration, in order to contain an even number of $Fe$ atoms with nearest-neighbor spins, placed at 4.483$\dot{A}$ and oriented in opposite directions.  
\begin{figure}
 \begin{center}
    \includegraphics[scale=0.7]{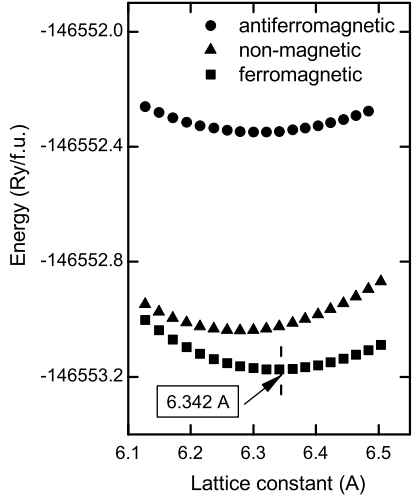} 
      \end{center}
   \caption{The calculated total energy in antiferromagnetic, paramagnetic and ferromagnetic states, as function of the lattice constant for $Ti_{2}FeSn$}
    \label{fig:optimizareTi2FeSn2cel}
\end{figure}
The minimum of the total energy for spin polarized configuration is lower than that for  non-magnetic and antiferromagnetic cases, therefore the ferromagnetic state is further used to analyze the $Ti_{2}FeSn$ compound. The equilibrium lattice constant obtained for spin-polarized setup is 6.342 $\dot{A}$ (see Fig. \ref{fig:optimizareTi2FeSn2cel}). 
\begin{figure}
 \begin{center}
    \includegraphics[scale=0.8]{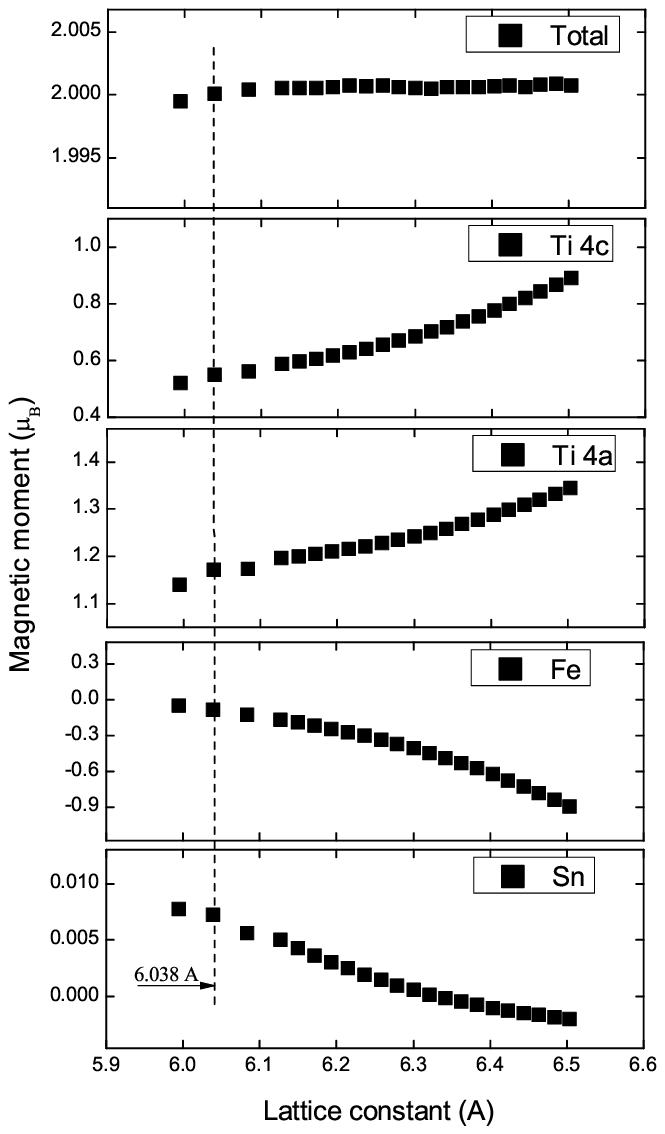} 
      \end{center}
   \caption{The total, $Ti$, $Fe$ and $Sn$ magnetic moments as function of lattice constant for $Ti_{2}FeSn$ compound. }
    \label{fig:magneticmomentTi2FeSn}
\end{figure}
 The calculated total spin magnetic moment and site resolved magnetic moments as function of lattice parameter are shown in Fig. \ref{fig:magneticmomentTi2FeSn}. 

The calculated total magnetic moment obtained is 2 $\mu_{B}$  and decreases when the lattice parameter decreases below 6.038 $\dot{A}$. The magnetic moment follows the new Slater-Pauling rule for the half-metallic Ti-based, full-Heusler alloys, with inverse Heusler structure, recently reported by \cite{Wei2012,Zheng2012}: $M_{t}=Z_{t}-18$ where $M_{t}$ is the total spin magnetic moment per unit cell, $Z_{t}$, the total number of valence electrons and 18 represents the number of occupied states in the spin bands. In terms of two-orbital two-electron stabilizing interactions, within the framework of density functional theory, the states from spin channels are occupied according to several aspects concerning ionic arguments, crystal structure of primitive cell, lattice parameter,  approximations made for the exchange and correlation interaction, energy threshold set between the core and valence states and also Brillouin zone integration mesh. Based on  ionic arguments, the most electropositive element transfers the valence  electrons to the more electronegative element. The purpose is to obtain stable closed shell ions. Since $Ti$ has 1.54 Pauling electronegativity, its valence atomic orbitals, formally should be depleted to fill the states of the more electronegative elements, i.e. $Fe$ and $Sn$ (1.83 and 1.96 Pauling units).  Strongly dependent by the atomic arrangement of atoms and environment, hybridization occurs whenever the sum of metallic radii (12-coordinated) of two first-neighbors exceeds the interatomic distance. Considering aforementioned description of chemical bondings, the number of uncompensated electron spins is 2, which gives the calculated total magnetic moment of $Ti_{2}FeSn$ compound. 

 The site resolved magnetic moments calculated at the equilibrium lattice parameter are 0.718 and 1.258 $\mu_{B}$ for $Ti$4c and $Ti$4a atoms while -0.486 and -0.0014 $\mu_{B}$ for Fe and Sn, respectively. In the interstitial region, the resulted magnetic moment is 0.509 eV. 
 
Similar results were found for the $Ti_{2}Fe$-based compounds \cite{KervanS2011,Wei2012,Zheng2012}. Nevertheless, the different bonding interactions between first-neighbors play a crucial role in determining the side magnetic moments of Ti atoms. For example, in the  semiconductor $Fe_{2}TiSn$, the magnetic moment of Ti atoms is 0 $\mu_{B}$, while in the half metallic full-Heusler $Fe_{2}TiSb$ is -0.46 $\mu_{B}$ \cite{Luo2012}. 
  
An inspection of the atom resolved magnetic moments reveals that the magnetic moments of $Ti$ atoms from both positions increase as the lattice expands. In spite of $Ti$ element by itself isn't magnetic, in the compound under study here, $Ti$ atoms from both positions, possess high spin magnetic moments at the optimized lattice parameter, being antiferromagnetically coupled with $Fe$ atoms. These results lead to the conclusion, that a $Ti_{2}Fe$ Heusler compound with $Sn$ yields a well-ordered and stable half-metallic ferrimagnet, similar to all
full-Heusler compounds, known as $Ti_{2}Fe$ -based reported  and described as ferrimagnet half-metals \cite{KervanN2011,Wei2012,Zheng2012,Ahmadian2012}. Other $Sn$ containing full Heusler compound, $Ti_{2}$-based, which has $Co$ atoms instead of $Fe$ as $Y$ element, also present a band gap in the minority spin channel, but it is  half-metallic ferromagnet($Ti$ and $Co$ atoms are ferromagnetically coupled), despite the nature of the first neighboring shells of $Y$ element are similar\cite{Birsan2012}.  
 
\begin{figure}
 \begin{center}
    \includegraphics[scale=0.8]{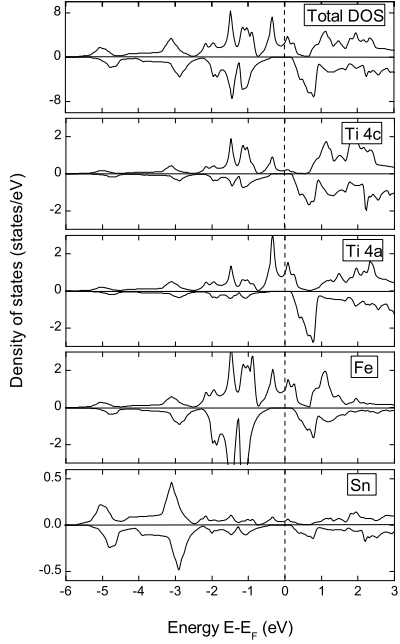} 
      \end{center}
   \caption{The spin-resolved total densities of states (DOS) and partial DOSs calculated at equilibrium lattice constant.}
    \label{fig:totaldosTi2FeSn}
\end{figure}

\begin{figure}
 \begin{center}
    \includegraphics[scale=0.6]{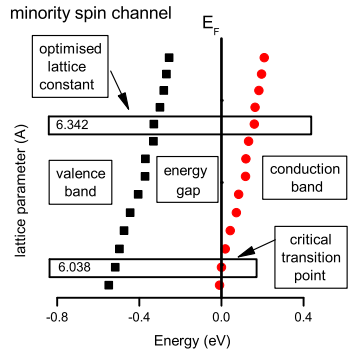} 
      \end{center}
   \caption{The positions of highest occupied states from the valence band (solid black squares) and of lowest unoccupied states from conduction band (solid red circles) of total DOSs for $Ti_{2}FeSn$ as function of lattice parameter.}
    \label{fig:gapTi2FeSn}
\end{figure}

The calculated total and partial density of states performed at equilibrium lattice constant are shown in Fig. \ref{fig:totaldosTi2FeSn}. The majority spin channel exhibits a typical metallic behavior, while the minority spin channel, a semiconducting character with a band gap around Fermi level, leading to a $100\%$  spin polarization. In both spin channels, the significant contribution to energy between -5.5 eV and -2.5 eV  comes not only from  $p$ electrons of Sn but also from $d$ electrons of Fe atoms which have a substantial contribution also between  -2.5 eV and -0.8 eV.
As Fig. \ref{fig:gapTi2FeSn}  shows, the size of the gap almost no changes but the Fermi level is shifted from the conduction band minimum to the center of the gap, while the lattice parameter increases. The transition from half-metallic to metallic character, illustrated in Fig. \ref{fig:gapTi2FeSn} and called critical transition point, occurs at 6.038 $\dot{A}$ when Fermi level is located at the bottom of conduction band. 

\begin{figure}
 \begin{center}
    \includegraphics[scale=0.7]{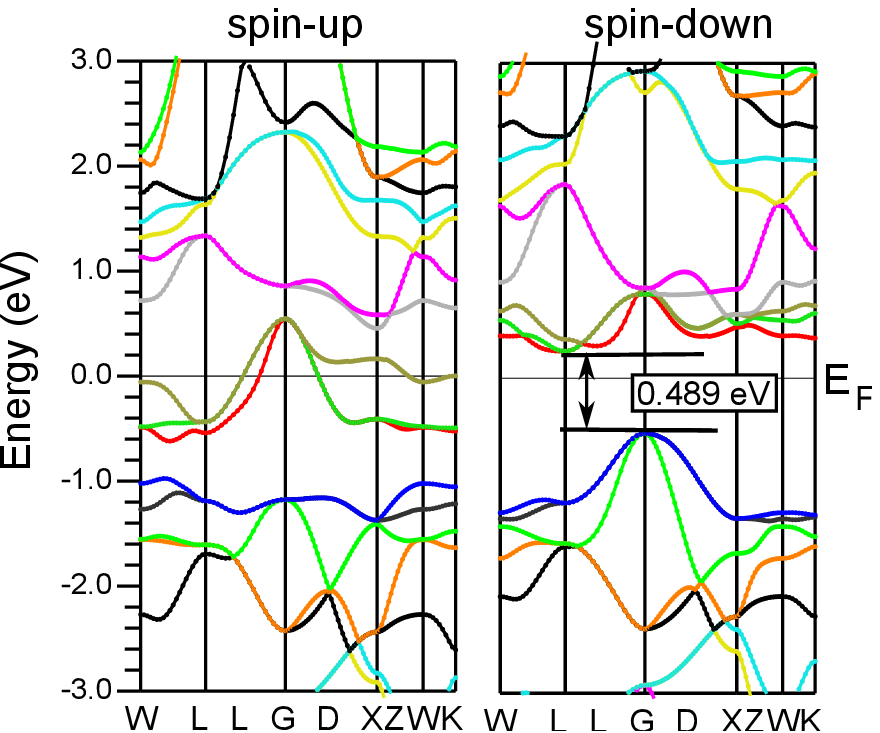} 
      \end{center}
   \caption{The band structure of $Ti_{2}FeSn$ for majority spin channel (spin-up band -left panel) and minority spin channel (spin-down band - right panel) electrons for optimized structure.}
    \label{fig:bandTi2FeSn}
\end{figure}

Fig. \ref{fig:bandTi2FeSn} displays the bandstructures of $Ti_{2}FeSn$ compound illustrating the metallic behavior from majority spin channel (spin-up band) and the semiconducting character from minority spin channel (spin-down band) with the indirect band gap around the Fermi level. At the equilibrium lattice constant the band gap from minority spin channel given by calculations is 0.489 eV. This value is determined by the difference in energy between the top of the valence band located at 0.329 eV below the Fermi level, at $\Gamma$ - point, and the bottom of the conduction band, placed at 0.160 eV above the $E_{F}$ at $L$ point. 

The band gap is formed not only as result of $Ti$(4a) - $Ti$(4c) hybridization but also due to $Ti$(4a) - Fe interaction. The double degenerated $d_{eg}$ orbitals of $Ti$(4a) atoms with energy values in the region corresponding to the anti-bonding states of the conduction band, couple with $d_{eg}$ orbitals of $Ti$(4c)and $Fe$ atoms, from valence band, corresponding to the bonding states. The interaction between $Ti$(4a) and $Fe$ is strong as well, although it is determined by the antiferromagnetic coupling between atoms. Additionally, it should be mentioned that the shortest distances from the unit cell are $Ti$(4a)-$Ti$(4c) and $Ti$(4c)-Fe, with close values near to 2.745 $\dot{A}$. The origin of the band gap in the compound under study here is similar to the band gap calculated for $Ti_{2}YAl$  with ($Y=V, Cr,Mn,Fe,Co,Ni,Cu,and\,Zn$) \cite{Zheng2012}, which reported that the semiconducting gap appears as result of hybridization of the $d$ states between $Ti$(4a)-$Ti$(4c) coupling and $Y$ atoms,  but somewhat different than the origin of the minority spin channel band gap from $Ti_{2}YZ$ ($Y = Fe, Co, Ni, Z = Al, Ga, In$) \cite{Wei2012} where, only the covalent hybridization between low valent and high valent atoms leads to the band gap formation. This result mismatch may be due to the different calculation method used or the approximations considered. The main group  element  has a crucial role to influence the size of the band gap and the structural stability. Hence, for $Ti_{2}FeSn$, the energy gap from minority spin channel is larger than the one for $Ti_{2}FeSi$ (0.45 eV \cite{KervanN2011} and smaller than the semiconducting gap of $Ti_{2}FeGe$ (0.84 eV \cite{Ahmadian2012}). Comparing with the   $Ti_{2}Fe$ compounds containing main group elements with less p- states than $Sn$ (i.e $Al$, $Ga$ or $In$ \cite{Wei2012}), the band gap of $Ti_{2}FeSn$ is smaller. With respect to $Ti_{2}FeAl$, further studies have to be done because two distinct gap values were reported: 0.53 eV \cite{Wei2012} and 0.33 eV \cite{Zheng2012}.

\section{Conclusions}
A novel Heusler compound, $Ti_{2}FeSn$ has been studied by using first-principles FPLAPW calculation method. The analysis of electronic structure and magnetic properties in the bulk, confirms that $Ti_{2}FeSn$ follows the new Slater-Pauling rule for the half-ferromagnetic full Heusler alloys, with inverse Heusler structure. The calculated total magnetic moment found is 2 $\mu_{B}$ and decreases for a lattice parameter lower than 6.038 $\dot{A}$. The majority spin channel presents metallic character, while the minority spin channel is semiconducting with a band gap of 0.489 eV,  at the equilibrium lattice constant 6.342 $\dot{A}$.  Overall,  the relative low magnetic moment and high spin polarization make $Ti_{2}FeSn$ Heusler compound attractive in spin torque transfer magnetic random access memory devices. 

\section{ACKNOWLEDGMENTS}
This work was financially supported from the projects  PNII IDEI 75/2011 and Core Program PN09-450103 of the Romanian Ministry of Education Research, Youth and Sport. 

\bibliographystyle{elsarticle-harv}

\end{document}